\documentclass[letterpaper,11pt]{article}

\pdfoutput=1 

\usepackage{jcappub} 

\usepackage[T1]{fontenc} 

\def\beq{\begin{equation}}
\def\eeq{\end{equation}}

\title{Solution for cosmological observables in the Starobinsky model of inflation}

\author[]{Gabriel Germ\'an,}
\author[]{Juan Carlos Hidalgo,}
\author[]{Luis E.~Padilla}

\affiliation[]{Instituto de Ciencias F\'{\i}sicas, Universidad Nacional
Aut\'onoma de M\'exico,\\ Av. Universidad s/n, Cuernavaca, Morelos, 62210, Mexico}

\emailAdd{gabriel@icf.unam.mx}
\emailAdd{hidalgo@icf.unam.mx}
\emailAdd{lepadilla@icf.unam.mx}

\abstract{This paper focuses on the Starobinsky model of  inflation in the Einstein frame. We derive solutions for various cosmological observables, such as the scalar spectral index $n_s$, the tensor-to-scalar ratio $r$ and their runnings, as well as the number of $e$-folds of inflation, reheating, and radiation, with minimal assumptions. The impact of reheating on inflation is explored by constraining the equation of state parameter  $\omega_{re}$ at the end of reheating. An equation linking inflation with reheating is established, which is solved for the spectral index $n_s$. Using consistency relations of the model, we determine the other observables while the number of $e$-folds during inflation $N_k$, and the number of $e$-folds during reheating $N_{re}$ are determined by their respective formulas involving $n_s$. We find remarkable agreement between the Starobinsky model and current measurements of the power spectrum of primordial curvature perturbations and the present bounds on the spectrum of primordial gravitational waves.
}
\begin{document}
\maketitle
\flushbottom
\section{Introduction}\label{int}
The Starobinsky model of inflation, proposed by A. Starobinsky in 1980 \cite{Starobinsky:1980te}, is a geometric model that incorporates linear and quadratic terms of the scalar curvature in its action, distinguishing it from other inflationary models (for reviews on inflation, see e.g., \cite{Linde:1984ir}-\cite{Martin:2018ycu}). By expressing the action in the Einstein frame, a scalar field potential emerges, aligning the model with typical inflation models. However, despite being proposed more than 40 years ago, most of the literature has focused on determining ranges for cosmological quantities like the scalar spectral index $n_s$, tensor-to-scalar ratio $r$, and the number of e-folds during inflation, denoted by $N_k$. In this study, we adopt an approach that allows us to derive solutions for the observables, as well as the number of e-folds during inflation, reheating and radiation, with minimal assumptions. We impose reheating conditions on inflation and obtain equations that enable us to solve for the desired quantities (for reviews on reheating, see e.g., [6]-[8]).
To understand the impact of reheating on inflation, we constrain the equation of state parameter (EoS) at the end of reheating, denoted as $\omega_{re}$. By establishing a connection between inflation and reheating, we derive an equation that determines the scalar spectral index, $n_s$. Using consistency relations within the model, we determine the remaining observables. Furthermore, we calculate the number of e-folds during inflation, reheating, and radiation, denoted as $N_k$, $N_{re}$, and $N_{rd}$, respectively.

The Starobinsky model modifies Einstein's theory of general relativity by introducing additional terms in the action. The action consists of the Einstein-Hilbert term, which is proportional to the Ricci scalar $R$, and an extra term proportional to the square of the Ricci scalar, $R^2$. The original Starobinsky model's action is given by 
\begin{equation}
\label{original}
S =\int d^4 x \sqrt{-g}\left[\frac{M_{Pl}^2}{2}\left(R + \frac{1}{6M^2}R^2\right) + L_m \right], 
\end{equation}
where  $g$ is the determinant of the metric tensor,  and $M_{Pl}=2.44\times 10^{18} \,\mathrm{GeV}$ is the reduced Planck mass. The parameter $M$ is related to the energy scale of inflation, and $L_m$ represents the Lagrangian density for matter fields. 
The inclusion of the $R^2$ term leads to modified equations of motion and a modified theory of gravity. During inflation, the $R^2$ term dominates over the Einstein-Hilbert term, resulting in exponential expansion. The $R^2$ term introduces a repulsive interaction that counteracts the attractive gravitational behavior, driving the accelerated expansion of the universe. This unique feature of the Starobinsky model provides a mechanism for inflation based solely on modifications to the gravitational sector, without the need for additional scalar fields.
The dominance of the $R^2$ term during inflation has implications for the dynamics of gravity and the resulting accelerated expansion. It allows for a prolonged period of inflation, resolving significant cosmological puzzles such as the horizon problem and the flatness problem. This extended exponential expansion is responsible for the observed large-scale homogeneity and isotropy of the universe.

The model, originally defined in the Jordan frame, is equivalent to a single-field model with an asymptotically flat potential when transformed through a conformal transformation to the Einstein frame. Additionally, by considering the Standard Model fields as minimally coupled to gravity in the Jordan frame, the transformation to the Einstein frame induces a coupling between these fields and the inflaton, which provides a natural mechanism for graceful exit and reheating~\cite{Vilenkin:1985md}, \cite{Faulkner:2006ub}, \cite{Gorbunov:2010bn}.

The model in the Einstein frame is obtained through a conformal transformation applied to the metric. This transformation is given by
\begin{equation}
g_{\mu\nu}\rightarrow e^{\sqrt{\frac{2}{3}}\frac{\phi}{M_{Pl}}}g_{\mu\nu}.
\end{equation}
Applying this transformation to the Starobinsky model, we obtain the action
\begin{equation}
S=\int d^{4}x \sqrt{-g}\left(\frac{M_{Pl}^{2}}{2}R-\frac{1}{2}\partial_{\mu}\phi\partial^{\mu}\phi-V(\phi)\right).
\end{equation}
Within this formulation, the potential of the scalar field takes the following form
\begin{equation}
\label{Spot}
V(\phi)=V_{0}\left(1-e^{-\sqrt{\frac{2}{3}}\frac{\phi}{M_{Pl}}}\right)^{2},
\end{equation}
where $V_{0}$ is defined as $\frac{3}{4}M_{Pl}^{2}M^2$.
By having the model expressed in the Einstein frame and identifying a scalar potential $V(\phi)$, we can employ the usual expressions for single-field inflation under the slow-roll approximation.

The Starobinsky model is remarkably consistent with current measurements of the power spectrum of primordial curvature fluctuations and the constraints on primordial gravitational waves. These measurements are derived from collaborations such as Planck and BICEP/Keck~\cite{Akrami:2018odb}, \cite{BICEP:2021xfz}, \cite{Tristram:2021tvh}. We consider the bounds provided by the Table 3 of \cite{Akrami:2018odb} for the cosmological model $\Lambda$CDM$+r+dn_s/d\ln k$ with the data set Planck TT,TE,EE+lowE+lensing +BK15+BAO. These bounds offer constraints on the parameters and observables within the specific cosmological model and data combination
\begin{equation}
n_s =0.9658\pm 0.0040    \quad    (68\%\,\, C.L.),
\label{boundsns} 
\end{equation}
\begin{equation}
r < 0.068     \quad    (95\%\,\, C.L.).
\label{boundcr} 
\end{equation}

The Starobinsky model predicts specific patterns in the anisotropies of the cosmic microwave background (CMB), including a characteristic damping scale in the power spectrum. Future observations from experiments such as the Simons Observatory \cite{simons} and the CMB-S4 \cite{cmbs4} collaboration are anticipated to yield more accurate measurements of the CMB. The Simons Observatory comprises four telescopes positioned at an elevation of 5200 meters in the Atacama Desert of Chile, while the CMB-S4 collaboration aims to enhance our understanding of key cosmological parameters by conducting precise measurements of the CMB's temperature and polarization. Specifically, the improved measurements of the CMB are expected to provide additional constraints on the Starobinsky model and other inflationary models, refining our understanding of the early universe and its evolution. These experiments aim to study the CMB with high sensitivity and precision using advanced detectors, larger arrays, multiple frequencies, improved angular resolution, and careful site selection. These experiments will provide valuable data to refine our understanding of the early universe, test inflationary models like the Starobinsky model, and investigate fundamental physics.

The paper is organized as follows: Section \ref{re} introduces the concept of reheating and establishes a general equation that connects inflation with reheating. This equation is then used to solve for $n_s$ in the context of the Starobinsky model, as presented in Section \ref{model}. In this section, we define the model and provide definitions for relevant quantities. Additionally, we derive an expression for the number of $e$-folds during inflation in terms of the spectral index $n_s$. Furthermore, we employ the consistency relations of the model to determine other observables. The analysis also encompasses the calculation of the number of $e$-folds during inflation, reheating, and the radiation era. Finally, Section \ref{con} provides the conclusion of our paper.
\section{Reheating constraints}\label{re}
Models of inflation can be related to cosmological observables, which, to first order in the slow-roll (SR) approximation, are expressed as (see, for example, \cite{Lyth:1998xn} and \cite{Liddle:1994dx})
\begin{eqnarray}
n_{t} &=&-2\epsilon = -\frac{r}{8} , \label{Int} \\
n_{s} &=&1+2\eta -6\epsilon ,  \label{Ins} \\
n_{sk}\equiv \frac{d n_s}{d \ln k} &=&16\epsilon \eta -24\epsilon ^{2}-2\xi_2, \label{Insk} \\
n_{tk}\equiv \frac{d n_t}{d \ln k} &=&4\epsilon\left( \eta -2\epsilon\right), \label{Intk} \\
A_s &=&\frac{1}{24\pi ^{2}} \frac{V}{\epsilon\, M_{Pl}^{4}}. \label{IA} 
\end{eqnarray}
Here, $M_{Pl}=2.43568\times 10^{18} \mathrm{GeV}$ is the reduced Planck mass, $r$ denotes the tensor-to-scalar ratio, $n_s$ represents the scalar spectral index, $n_{sk}$ its running (which is commonly denoted as $\alpha$), $n_t$ represents the tensor spectral index, and $n_{tk}$ represents its running, in a self-explanatory notation. The amplitude of density perturbations at a particular wave number $k$ is denoted by $A_s$. All quantities are evaluated at the moment of horizon crossing at wavenumber $k=0.05$/Mpc. The SR parameters involved in the above expressions are
\begin{equation}
\epsilon \equiv \frac{M_{Pl}^{2}}{2}\left( \frac{V^{\prime }}{V }\right) ^{2},\quad
\eta \equiv M_{Pl}^{2}\frac{V^{\prime \prime }}{V}, \quad
\xi_2 \equiv M_{Pl}^{4}\frac{V^{\prime }V^{\prime \prime \prime }}{V^{2}},
\label{Spa}
\end{equation}
where primes on $V$ denote derivatives with respect to the inflaton $\phi$.

Expanding on earlier work \cite{Liddle:2003as, Dodelson:2003vq, Liddle:1994dx}, it is possible to derive an equation for the number of $e$-folds during reheating \cite{Dai:2014jja, Munoz:2014eqa} by relating the comoving Hubble scale wavenumber $k$ at horizon crossing to the present scale wavenumber $k_0=a_0 H_0$ as follows (also see \cite{German:2020cbw}, \cite{German:2020iwg} for further details)
\begin{equation}
\label{Nre1}
N_{re}= \frac{4}{1-3\, \omega_{re}}\left(-N_{k}-\frac{1}{3} \ln\left[\frac{11 g_{s,re}}{43}\right]-\frac{1}{4} \ln\left[\frac{30}{\pi^2 g_{re} } \right] -\ln\left[\frac{ k}{a_0 T_0} \right]-\frac{1}{4}\ln\left[\frac{\rho_e }{H_k^4} \right]\right).
\end{equation}
In the above equation, the number of degrees of freedom of species at the end of reheating is denoted by $g_{re}$, while $g_{s,re}$ represents the entropy number of degrees of freedom after reheating. The energy density at the end of inflation is denoted by $\rho_{e}$, with $a_0$ and $T_0$ representing the scale factor and temperature today, respectively.
The energy density above is model-dependent and can be expressed as $\rho_e=\frac{3}{2}V_e$. Here, $V_e$ represents the potential of the model at the end of inflation, while $H_k$ is the Hubble function at the comoving Hubble scale wavenumber $k$. 

An expression for the number of $e$-folds during reheating, in terms of energy densities, can be obtained \cite{Dai:2014jja}  by solving the fluid equation assuming a constant EoS $\omega_{re}$  
\begin{equation}
\label{Nre2}
N_{re}= \frac{1}{3(1+\omega_{re})}\ln\left[\frac{\rho_e}{\rho_{re}}\right]= \frac{1}{3(1+\omega_{re})}\ln\left[\frac{\frac{3}{2}V_e}{\frac{\pi^2g_{re}}{30}T_{re}^4}\right],
\end{equation}
where $T_{re}$ is the reheating temperature. 
From Eqs.~(\ref{Nre1}) and (\ref{Nre2})  we get
\begin{equation}
\label{Nk4}
N_{k}=- \frac{1-3\, \omega_{re}}{12(1+\omega_{re})}\ln[\frac{45V_e}{\pi^2g_{re}T_{re}^4}]-\frac{1}{3} \ln\left[\frac{11 g_{s,re}}{43}\right]-\frac{1}{4} \ln\left[\frac{30}{\pi^2 g_{re} } \right] -\ln\left[\frac{ k}{a_0 T_0} \right]-\frac{1}{4}\ln\left[\frac{27V_e M_{Pl}^4 }{2V_k^2} \right],
\end{equation}
where $V_k\equiv V(\phi_k)$ is the potential  at the comoving Hubble scale wavenumber $k$. We can express the potential as $V(\phi)=V_0 f(\phi)$, where $V_0$ represents the overall scale and $f(\phi)$ contains all the terms of the potential that depend on $\phi$. This choice does not introduce any loss of generality. Eq. (\ref{Nk4}) simplifies as follows
\begin{equation}
\label{Nk4a}
N_{k}=\ln\left(\frac{\left(\frac{43}{11g_{s,re}}\right)^{1/3}g_{re}^{1/4}\sqrt{\pi}a_0T_0}{3\times 5^{1/4}k_p}\left(\frac{\pi^2g_{re}T_{re}^4}{45M_{Pl}^4}\right)^{\frac{1-3\omega_{re}}{12(1+\omega_{re})}}\left(\frac{V_0}{M_{Pl}^4}\right)^{\frac{1+3\omega_{re}}{6(1+\omega_{re})}}\frac{f^{1/2}(\phi_k)}{f^{\frac{1}{3(1+\omega_{re})}}(\phi_e)}\right).
\end{equation}
We can further simplify Eq.~(\ref{Nk4a}) eliminating $V_0$. By using Eq.~(\ref{IA}) we can solve for $V_0$ in terms of $\phi_k$ and the amplitude of scalar perturbations $A_s$ at horizon crossing
\begin{equation}
\label{V0}
V_0=\frac{12A_s\pi^2\left(f^{\prime}(\phi_k)M_{Pl}\right)^2}{f(\phi_k)^3}M_{Pl}^4 ,
\end{equation}
where $f^{\prime}(\phi_k)$ is the derivative of $f(\phi)$ with respect to $\phi$ evaluated at $\phi=\phi_k$. Finally, Eq. (\ref{Nk4a}) can be written as
\begin{equation}
\label{Nk4b}
N_{k}=T_1+\frac{1}{3(1+\omega_{re})}\ln\left(\frac{\left(f^{\prime}(\phi_k)M_{Pl}\right)^{1+3\omega_{re}}}{f(\phi_k)^{3 \omega_{re}}f(\phi_e)}\right),
\end{equation}
where the term $T_1$ is given by
\begin{equation}
\label{T1}
T_1=\ln\left(\frac{\left(\frac{43}{11g_{s,re}}\right)^{1/3}g_{re}^{1/4}\sqrt{\pi}a_0T_0}{3\times 5^{1/4}k_p}\left(\frac{\pi^2g_{re}}{45}\right)^{\frac{1-3\omega_{re}}{12(1+\omega_{re})}}\left(12A_s\pi^2\right)^{\frac{1+3\omega_{re}}{6(1+\omega_{re})}}\right)+\frac{1-3\omega_{re}}{3(1+\omega_{re})}\ln \frac{T_{re}}{M_{Pl}}.
\end{equation}
We immediately notice that, for $\omega_{re}=1/3$, $T_1$  (and $N_k$) is independent of $T_{re}$. Eq. (\ref{Nk4b}) is a general equation connecting reheating and inflation, valid for any single field potential and any $\omega_{re}$. In the following section we show how to effectively use this approach by applying it to the Starobinsky potential.
\section{The Starobinsky model, observables and the number of $e$-folds}\label{model}

The potential for the Starobinsky model in the Einstein frame is given by
\begin{equation}
\label{Spot}
V(\phi)=V_{0}\left(1-e^{-\sqrt{\frac{2}{3}}\frac{\phi}{M_{Pl}}}\right)^{2}.
\end{equation}
By expressing the model in the Einstein frame and identifying a scalar potential $V(\phi)$, we can use standard expressions for slow-roll single-field inflation. This enables us to establish relationships between cosmological observables such as $r$, $n_{sk}$, $n_{tk}$, and the scalar spectral index $n_s$. Here, $n_{sk}$ represents the running of the spectral index, while $n_{tk}$ denotes the tensor running.

To find relations between $n_{s}$ and the rest of the quantities of interest, it is convenient to derive a closed-form expression for the inflaton field at horizon crossing, $\phi=\phi_k$. This can be accomplished by solving Eq.~(\ref{Ins}), which results in
\begin{equation}
\label{fik1}
\phi_{k}=\sqrt{\frac{3}{2}}M_{Pl}\ln \left(\frac{4+3\delta_{n_s}+4\sqrt{1+3\delta_{n_s}}}{3\delta_{n_s}}\right),
\end{equation}
where $ \delta_{n_s}\equiv 1-n_s.$
We can determine the number of $e$-folds during inflation $N_k$, after the pivot scale of wavenumber $k\equiv a_k H_k$ left the horizon, using the SR approximation
\begin{equation}
\label{Nk5}
 N_{k}=-\frac{1}{M_{Pl}^{2}}\int _{\phi_{k}}^{\phi_{e}} \frac{V}{V'} d\phi= \frac{3}{4}\left(e^{\sqrt{\frac{2}{3}}\frac{\phi_{k}}{M_{Pl}}}-e^{\sqrt{\frac{2}{3}}\frac{\phi_{e}}{M_{Pl}}}-\sqrt{\frac{2}{3}}\left(\frac{\phi_{k}}{M_{Pl}}-\frac{\phi_{e}}{M_{Pl}}\right)\right).
\end{equation}
Where $\phi_{e}$ is the field evaluated at the end of inflation which, following \cite{Ellis:2015pla}, we approximate as $\phi_e \approx 0.615 M_{Pl}$. Given the horizon exit value for $\phi_k$ by Eq.~(\ref{fik1}), it is possible to express $N_k$ in terms of the spectral index $n_s$.

\begin{figure}[ht!]
\centering
\includegraphics[trim = 0mm  0mm 1mm 1mm, clip, width=9.5cm, height=6.5cm]{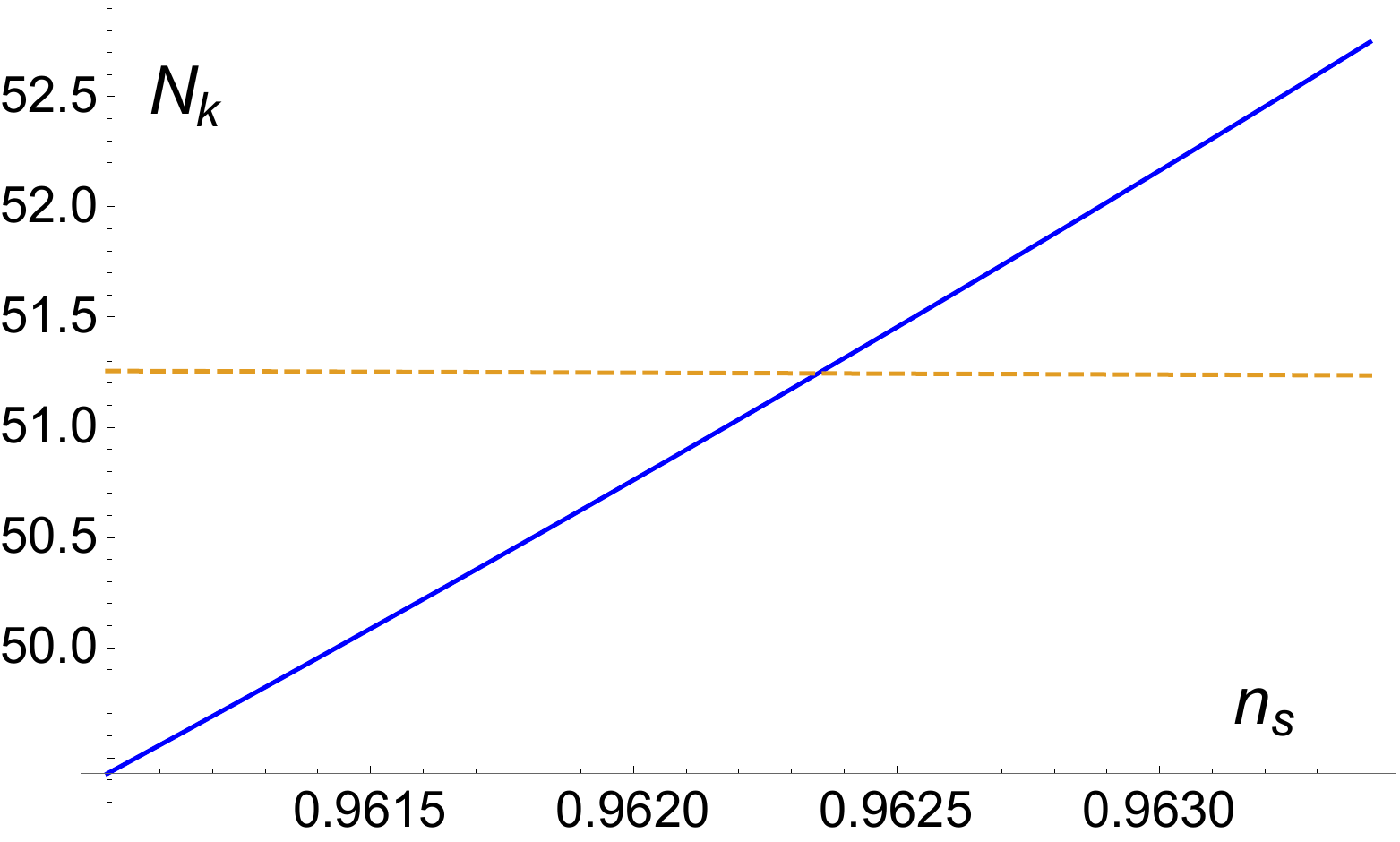}
\caption{The solid (blue) curve represents the lhs of Eq.~(\ref{Nk6}) while the dashed (orange) curve represents the slowly varying function of the rhs of the same equation. For the Starobinsky model in the Einstein frame defined by Eq.~(\ref{Spot}) we approximate the EoS by  $\omega_{re}=0$, and $T_{re}=3.1\times 10^9 \mathrm{GeV}$ \cite{Gorbunov:2010bn}.
The intersection point gives us the value $n_s=0.96235$ for the scalar spectral index and, from the consistency relations for the Starobinsky model given by Eqs.~(\ref{Int}), (\ref{r1}), (\ref{nsk1}), and (\ref{ntk1}), we obtain the other observables.}
	\label{Nks}
\end{figure}

At the origen the Starobinsky model is well approximated by a quadratic potential, in this case $\omega_{re}=0$. Also, for the Starobinsky model $T_{re}$ has been determined to be $3.1\times 10^9 \mathrm{GeV}$ \cite{Gorbunov:2010bn}. In this case $T_1\approx 58.7261+
 \frac{1}{3}\ln\frac{T_{re}}{M_{Pl}}\approx 51.8988$. From Eq.~(\ref{Nk4b}) we get
\begin{equation}
\label{Nk6}
N_{k}=T_1+\frac{1}{3}\ln\left(\frac{f^{\prime}(\phi_k)M_{Pl}}{f(\phi_e)}\right),
\end{equation}
where $f^{\prime}(\phi_k)=\frac{2}{M_{Pl}}\sqrt{\frac{2}{3}}e^{-\sqrt{\frac{2}{3}}\frac{\phi_k}{M_{Pl}}}\left(1-e^{-\sqrt{\frac{2}{3}}\frac{\phi_k}{M_{Pl}}}\right)$ and, at the end of inflation,  $f(\phi_e)= \left(1-e^{-\sqrt{\frac{2}{3}}\frac{\phi_e}{M_{Pl}}}\right)^{2}$. Having obtained $N_k$ in terms of $\phi_k$, and $\phi_k$ in terms on $n_s$, we can solve Eq.(\ref{Nk6}) directly for $n_s$ (see Fig.~\ref{Nks}). This yields the following result
\begin{equation}
\label{ns}
n_s= 0.96235,
\end{equation}
with the last digit rounded off.  

The consistency relations for the Starobinsky model \cite{Garcia:2023tkk} provide the following expressions for the tensor-to-scalar ratio $r$, the running of the spectral index $n_{sk}$, and the running of the tensor $n_{tk}$, in terms of $n_s$
\begin{equation}
\label{r1}
r=\frac{4}{3}\left(2-2\sqrt{1+3\delta_{n_s}}+3\delta_{n_s}\right),       
\end{equation}
\begin{equation}
\label{nsk1}
n_{sk}=\frac{1}{18}\left(4+9\delta_{n_s}-9\delta_{n_s}^2-\sqrt{1+3\delta_{n_s}}\left(4+3\delta_{n_s}\right)\right),
\end{equation}
\begin{equation}
\label{ntk1}
n_{tk}=\frac{1}{36}\left(8+3(4-3\delta_{n_s})\delta_{n_s}-8\sqrt{1+3\delta_{n_s}}\right),
\end{equation}
where $\delta_{n_s}\equiv 1-n_s$, as before. Using these equations, we can compute the values for the other observables. We can also determine the number of $e$-folds of inflation, reheating and, from entropy conservation after reheating \cite{German:2022sjd}, the number of $e$-folds of radiation
\begin{equation}
\label{Nrd}
N_{rd}\equiv \ln\left(\frac{a_{eq}}{a_r}\right)=\ln\left(\frac{a_{eq} T_{re}}{\left(\frac{43}{11 g_{s,re}} \right)^{1/3}a_0T_0}\right),
\end{equation}
where $a_{r}$ denotes the scale factor at the end of reheating or, equivalently, at the beginning of the radiation epoch and $a_{eq}$ is the scale factor at radiation-matter equality. Values of observables as well as number of $e$-folds are given in the Table~\ref{tabla}. 

Finally, we can use the expression 
\begin{equation}
N_{keq}\equiv \ln\left(\frac{a_{eq}}{a_k}\right)= \ln\left(\frac{a_{e}}{a_k}\right)+ \ln\left(\frac{a_r}{a_e}\right)+ \ln\left(\frac{a_{eq}}{a_r}\right)=N_{k}+N_{re}+N_{rd},
\end{equation}
as a consistency check. This equation can be written more concisely as $N_{keq}=\ln\left(\frac{a_{eq} H_k}{k_p}\right)=\ln \left(\frac{a_{eq}\pi\sqrt{A_s r}}{\sqrt{2}k_p}\right)$ \cite{German:2020iwg}. Thus, the number of $e$-folds from the time scales of wavenumber $k=a_k H_k$ leave the horizon at $a_k$ to the time of radiation-matter equality at $a_{eq}$ is essentially given by the parameter $r$, equivalently, by the value of the scale factor at horizon crossing $a_k$ \cite{German:2023yer}.
We find that $N_{k}+N_{re}+N_{rd}\approx 113.182$, and the same value using the formula $N_{keq}=\ln \left(\frac{a_{eq}\pi\sqrt{A_s r}}{\sqrt{2}k_p}\right)$. 
\begin{table*}[htbp!]
 \begin{center}
{\begin{tabular}{cccc}
\small
Parameter & Value & Parameter & Value  \\
\hline \hline
{$k_p $} & $0.05Mpc^{-1}$ & $T_{re}$ & $3.1\times 10^{9}\mathrm{GeV}$ \\
{$T_0 $}   & $2.7255\,K$ & $\phi_{e}$ & $0.615 M_{Pl}$  \\
{$A_s $}  & $2.1\times 10^{-9}$ & {$a_{eq} $}& $2.9\times 10^{-4}$ \\
 \hline \hline
Observable & Value & $e$-folds & Value  \\
\hline \hline
{$n_{s}$} & $0.96235$ & $N_k$ & $51.2$  \\
{$r$} & $0.00403$ & $N_{re}$ & $18.0$ \\
{$n_{t}$} & $-0.00050$ &  $N_{rd}$ & 43.9\ \\
{$n_{sk}$} & $-0.00072$ & $N_k+N_{re}+N_{rd}$ & 113.2\\
{$n_{tk}$} & $-0.000019$ & $N_{keq}$ & $113.2$  \\
\hline\hline
\end{tabular}}
\caption{Above, we have listed the parameter values used in the calculations, while below, we present the values obtained for the observables and number of $e$-folds. Note that for $g_{re}\approx g_{s,re}$, Eq.~(\ref{Nk6}) is practically independent of both $g_{re}$ and $g_{s,re}$. We solve Eq.~(\ref{Nk6}) for the spectral index $n_s$, and by using the consistency relations given by (\ref{Int}), (\ref{r1}), (\ref{nsk1}), and (\ref{ntk1}), we obtain the other observables. The number of $e$-folds during inflation $N_k$, reheating $N_{re}$, and radiation $N_{rd}$ are calculated using the equations (\ref{Nk5}), (\ref{Nre2}), and (\ref{Nrd}), respectively. For an explanation about the presence of the term $N_{keq}$, see the last paragraph of Section \ref{model}. We observe that all the values obtained are within ranges calculated previously with a different procedure  \cite{Garcia:2023tkk} (see also  \cite{Gorbunov:2012ns}) and are in remarkable agreement with current measurements of the power spectrum of primordial curvature fluctuations and the present bounds on the spectrum of primordial gravitational waves.}
\label{tabla}
\end{center}
\end{table*}

\section{Conclusions}\label{con}
Our approach allows us to find solutions for the cosmological observables, including the number of $e$-folds during inflation, reheating, and radiation, with minimal assumptions. By imposing the reheating conditions, we establish a connection between inflation and reheating and derive Eq. (\ref{Nk4b}) which, for the Starobinsky model and $\omega_{re}=0$, is solved for the spectral index $n_s$. We use the consistency relations of the model to determine the values for the other observables. The number of $e$-folds during inflation $N_k$ and the number of $e$-folds during reheating $N_{re}$ are also determined by their respective formulas involving $n_s$, while the number of $e$-folds during radiation $N_{rd}$ is determined by the reheating temperature $T_{re}$. The results show remarkable agreement between the Starobinsky model and current measurements of the power spectrum of primordial curvature fluctuations and the present bounds on the spectrum of primordial gravitational waves.

\section*{Acknowledgments}

{The authors  acknowledge support from program UNAM-PAPIIT, grants IN107521 “Sector Oscuro y Agujeros Negros Primordiales” and IG102123 ``Laboratorio de Modelos y Datos (LAMOD) para proyectos de Investigaci\'on Cient\'ifica: Censos Astrof\'isicos".  L. E. P. and J. C. H. acknowledge sponsorship from CONAHCyT Network Project No.~304001 ``Estudio de campos escalares con aplicaciones en cosmolog\'ia y astrof\'isica'', and through grant CB-2016-282569.  The work of L. E. P. is also supported by the DGAPA-UNAM postdoctoral grants program, by CONAHCyT M\'exico under grants  A1-S-8742, 376127 and FORDECYT-PRONACES grant No. 490769.}
\\
\\
Data Availability Statement: No Data associated in the manuscript.

\end{document}